\documentclass[aps,twocolumn,longbibliography,
reprint,%
]{revtex4-1}

\usepackage{amsmath,natbib}
\usepackage{amssymb}
\usepackage{graphicx,color,tabto}% Include figure files
\usepackage{dcolumn}% Align table columns on decimal point
\usepackage{bm}% bold math

\allowdisplaybreaks[1]

\newcommand{\markred}[1]{\textcolor{black}{#1}}
\newcommand{\markblue}[1]{\textcolor{black}{#1}}

\begin{document}

\preprint{Preprint \#}

\title[Lattices of hydrodynamically interacting flapping swimmers]{Lattices of hydrodynamically interacting flapping swimmers}% Force line breaks with \\

\author{Anand U. Oza}%
\email{oza@njit.edu}
\affiliation{Department of Mathematical Sciences, New Jersey Institute of Technology, Newark, New Jersey 07102, USA}%

\author{Leif Ristroph}
  \affiliation{Courant Institute of Mathematical Sciences, New York University, New York, New York 10012, USA}%

\author{Michael J. Shelley}
\affiliation{Courant Institute of Mathematical Sciences, New York University, New York, New York 10012, USA}
\affiliation{Center for Computational Biology, Flatiron Institute, New York, New York 10010, USA}%
 
\date{\today}% It is always \today, but any date may be explicitly specified

\begin{abstract}
Fish schools and bird flocks exhibit complex collective dynamics whose self-organization principles are largely unknown. The influence of hydrodynamics on such collectives has been relatively unexplored theoretically, in part due to the difficulty in modeling the temporally long-lived hydrodynamic interactions between many dynamic bodies. We address this through a novel discrete-time dynamical system (iterated map) that describes the hydrodynamic interactions between flapping swimmers arranged in one- and two-dimensional lattice formations. Our 1D results exhibit good agreement with previously published experimental data, in particular predicting the bistability of schooling states and new instabilities that can be probed in experimental settings. For 2D lattices, we determine the formations for which swimmers optimally benefit from hydrodynamic interactions. \markblue{We thus obtain the following hierarchy: while a side-by-side single-row \lq\lq phalanx\rq\rq\, formation offers a small improvement over a solitary swimmer, 1D in-line and 2D rectangular lattice formations exhibit substantial improvements, with the 2D diamond lattice offering the largest hydrodynamic benefit.} Generally, our self-consistent modeling framework may be broadly applicable to active systems in which the collective dynamics is primarily driven by a fluid-mediated memory.
\end{abstract}

%\keywords{Flocking, schooling, flapping flight, nonlinear dynamics}%Use showkeys class option if keyword

\maketitle

\section{Introduction}
The complex collective dynamics of fish schools and bird flocks have long fascinated physicists, biologists and mathematicians~\cite{Pavlov_Review,Bajec_Review}. In addition to their biological relevance, they are living examples of active systems~\cite{CavagnaAnnRev,Marchetti_Review,Ramaswamy_Review,Saintillan_Rev} in which energy input by the individual constituents gives rise to organized collective phenomena. While there has been considerable experimental and theoretical progress in characterizing \lq\lq dry\rq\rq\, active systems (e.g. shaken granular rods~\cite{Narayan_Granular,Kudrolli_Rods}) % and disks ~\cite{Deseigne_Disks}) 
and the collective behavior of biological systems at the microscale (e.g. bacterial suspensions~\cite{Sokolov_viscosity,Goldstein_Turbulence,Zhang_Bacteria,Saintillan1}), significantly less is known about the role of hydrodynamic interactions in mediating schooling and flocking behavior in collectives of larger animals. More generally, the influence of inertial fluid flows and the consequent long-lived hydrodynamic interactions 
on collective behavior remains poorly understood. In contrast to the low Reynolds number (Stokes) regime in which microswimmers operate, fluid-mediated memory could significantly impact animal schools and also nonliving systems dominated by wave-particle interactions. As an example of the latter, oil droplets bouncing on a vertically vibrating fluid bath~\cite{annualReview} are known to exhibit crystal-like bound states~\cite{Lieber-lattices,lattices} % cut ref to lattice2 (Eddi EPL 2011)
as a result of their coupling through surface waves. We present here a modeling framework for the long-lived hydrodynamic interactions between swimmers, with a view to understanding how their collective dynamics might be mediated by flow-induced forces.

The influence of hydrodynamics on schooling and flocking behavior in biological systems has been the subject of intense debate in the scientific literature. While some analyses of starling cluster flock data~\cite{Bialek1,AttanasiNatPhys} % cut ref to MoraNatPhys
focus on the behavioral mechanisms behind flocking, a recent analysis~\cite{Portugal} of flying ibises in V-formations demonstrated coherence of the birds' wing tip paths, which enables upwash capture from their neighbors to be maximized. %but contradicts work on pigeons~\cite{Usherwood_Pidgeon}. These results are consistent with prior measurements on pelicans~\cite{Pelican1}, and theoretical analyses have demonstrated %reduced power requirements for weight support both drag reduction and enhanced lift generation when birds execute formation flight~\cite{Formation,Hummel,Badgerow_V,Maeng_Geese}. 
Individual fish have been shown to sense and respond to environmental hydrodynamic cues~\cite{Leif_LateralLine}, and benefit from external flows by harnessing the energy from vortices~\cite{Triantafyllou_Review,Liao_Review,Liao_Science}. % Cut ref to Liao_Karman
While some studies have argued against a hydrodynamic function for fish schools~\cite{Partridge} and instead focused on the social interactions between fish~\cite{Jolles1,Swain1,Tunstrom1}, a number of observations~\cite{Partridge2,Herskin,Marras,AshrafPNAS} have indicated that schooling fish could benefit from hydrodynamic interactions by realizing significant energy savings. 

Theoretical models of flocks and schools have also largely ignored hydrodynamic interactions, and instead have shown that self-organized collective behavior may emerge from a relatively simple set of behavioral interaction rules~\cite{Couzin1,LopezReview}. Examples are the seminal works of Huth \& Wissel~\cite{Huth} and Viscek {\it et al.}~\cite{Viscek}, which have been extended to other discrete-time models with more elaborate interaction rules~\cite{Barberis1}. % Cut ref to Chate_Viscek
%Coarse-graining the Viscek model gives the Toner-Tu continuum theory~\cite{TonerTu}, which has been generalized to incorporate inertial spin fluctuations~\cite{CavagnaPRL,Yang_Turning} and flocking on curved surfaces~\cite{MarchettiFlock}. 
Far-field hydrodynamic interactions have been recently incorporated into self-propelled particle models of swimmers subject to phenomenological behavioral rules~\cite{Gazzola_JFM,Filella_Hydro}, but such models neglect the vorticity-induced forces thought to be relevant for schooling fish~\cite{Liao_Review}. 

\markblue{Fish schools and bird flocks can exhibit orderly lattice-like formations, although field observations are diverse and sometimes contradictory~\cite{Pavlov_Review,Bajec_Review}. Prior observations revealed that fish schools may adopt lattice configurations in a statistical sense~\cite{Cullen}. A number of fish species (e.g. minnow, bream, saithe, herring) adopt schooling formations reminiscent of 3D tetrahedral and cubic lattices, but others (e.g. cod) adopt less ordered configurations~\cite{Partridge_3D,Pitcher_Minnow}. 
Relatively rigid school structures, which are treated in this paper, have also been observed. For example, there are a number of accounts of fish swimming in linear chains~\cite{Gudger}. %For instance, giant bluefin tuna commonly school in a parabolic configuration~\cite{Partridge2}. 
In their studies of red nose tetra fish, Ashraf {\it et al.}~\cite{AshrafPNAS,AshrafINT} noted a prevalence of 2D diamond lattice formations at low swimming speeds and \lq\lq phalanx\rq\rq\, formations at higher speeds, the latter being a side-by-side arrangement of swimmers roughly equispaced in a single line perpendicular to the swimming direction. These two schooling formations have also been observed in bluefin tuna~\cite{Newlands_tuna}, for which small schools tend to adopt a phalanx formation while larger schools adopt a diamond lattice. With respect to bird flocks, ibises have been observed to obtain aerodynamic and energetic benefits from in-line formation flight~\cite{Portugal}. Certain species (e.g. pelicans~\cite{Pelican1}, Canada geese~\cite{Badgerow_V}, ibises~\cite{Portugal}) are thought to benefit from their observed V-formation flight, while it is claimed that others (e.g. pigeons~\cite{Usherwood_Pidgeon}, pink-footed geese~\cite{Cutts_V}) do not. Corcoran \& Hedrick~\cite{Corcoran_V} have recently identified a new type of ordered configuration in shorebird flocks, the \lq\lq compound V-formation,\rq\rq\, wherein a given bird commonly flies roughly one wingspan to the side and to the back %between a half to one-and-a-half wingspans back 
from the bird in front of it. Such an ordered structure is observed at all spatial scales within an extended flock, unlike the purely local structure exhibited by starling cluster flocks~\cite{Bialek1,AttanasiNatPhys}.}

%Vortex-induced hydrodynamic interactions in fish schools were considered in the seminal work of Weihs~\cite{Weihs1,Weihs2}, who modeled a school as a 2D lattice of swimmers that flap in antiphase with respect to their lateral neighbors and shed point vortices. 
\markblue{In an attempt to explain such ordered structures, Weihs' seminal papers~\cite{Weihs1,Weihs2} considered} vortex-induced hydrodynamic interactions in a 2D lattice of swimmers. % were examined in the seminal works of Weihs~\cite{Weihs1,Weihs2}. 
By positing that fish seek to minimize their hydrodynamic drag while avoiding large flow velocity gradients, Weihs argued that a diamond lattice is the energetically optimal arrangement. While this model has been highly influential, it has not been developed further, due to the lack of experimental confirmation and various modeling assumptions. Specifically, the school's swimming dynamics was not accounted for; hence, the speed and efficiency were not self-consistently calculated, and there was no consideration of the stability of the most efficient state. Moreover, the influence of the streamwise spacing between swimmers was largely neglected, as the swimmers were assumed to be separated by at least five flapping wavelengths. %, so that the induced streamwise velocity is approximately periodic. 
These deficits are addressed through the model we present herein.

Various groups have conducted numerical simulations of the Navier-Stokes  
equations coupled to an immersed body's dynamics, and have thus studied flapping wings~\cite{Streitlien,Anderson_Foil,Alben_Flexible2,Moore_Flexible,Akhtar_Tandem}
%rigid~\cite{Streitlien,Anderson_Foil,Streitlien2, Andersen_Foil} and flexible wings~\cite{Alben_Flexible,Alben_Flexible2,Moore_Flexible}; 
%tandem wings~\cite{Akhtar_Tandem} and elastic filaments~\cite{Zhu_Flapping}; 
and deformable bodies with more realistic fish-like kinematics~\cite{Zhu_Flapping,MacIver_Fins,Tytell_Swim,Borazjani_Swim,Liu_CompVortex,Gazzola_JCP,Hieber_IB,Maertens_Optim}. %Cut refs to Tytell_Stiffness,Bhalla_Swim,Deng
Hemelrijk {\it et al.}~\cite{Hemelrijk} and Daghooghi {\it et al.}~\cite{Daghooghi} modeled a fish school by numerically simulating a swimmer with doubly-periodic boundary conditions in 2D and 3D, respectively, and found that swimmers move faster in schools than in isolation. However, neither study examined the dependence of the speed on the streamwise distance between swimmers. While these studies allowed for the complex flow structures around flapping bodies to be quantitatively studied, simulations of fish schools are computationally challenging because of the large Reynolds number of the associated flow and the number of interacting bodies, prohibiting a detailed parametric study of lattice formations.

Understanding the role of hydrodynamic interactions in fish schools may thus benefit from a simplified physical system amenable to theoretical analysis. An example is the recent experimental work of Becker {\it et al.}~\cite{Becker}, who realized an in-line %a linear 
formation of swimmers using freely-translating, periodically heaving wings in a cylindrical water tank. They observed that the system exhibited a bistability of \lq\lq schooling states\rq\rq\, and spontaneously locked into either a slow mode or fast mode, the latter of which exhibited a significant speed increase relative to an isolated swimmer. The experiments were extended by Ramananarivo {\it et al.}~\cite{Sophie} to allow tandem swimmers to dynamically select both their speeds and relative positions. 

We present here a modeling framework for understanding the hydrodynamic interactions among flapping swimmers at high Reynolds number. %, with a view to rationalizing the experimental results of Becker {\it et al.}~\cite{Becker} 
Our conceptually simple model is rich enough to incorporate the essential features of the swimmers' hydrodynamic interactions, while allowing for analytical determination of the model's exact solutions and their stability properties. Crucially, our model differs from other self-propelled particle models in that the swimmers' shed vortices are accounted for, so the thrust on a swimmer depends on the system's history. The swimmers thus interact through a fluid-mediated memory, stored through the collective shed vorticity, and we self-consistently solve for the formation's emergent speed. The model yields insight into experiments on interacting wings~\cite{Becker} and predicts instabilities that can be explored in the laboratory. We also extend our model to 2D and determine the lattice formations that allow for the greatest speed and efficiency, thus addressing the questions first posed by Weihs~\cite{Weihs1,Weihs2}. \markblue{Specifically, we obtain the following hierarchy: while a phalanx formation (\S\ref{Sec:Phalanx}) offers a small improvement over a solitary swimmer, 1D in-line (\S\ref{Sec:Line}) and 2D rectangular lattice formations (\S\ref{Sec:Rect}) exhibit substantial improvements, with the 2D diamond lattice (\S\ref{Sec:Diamond}) offering the largest hydrodynamic benefit.}

\section{Simple model of an in-line formation}\label{Sec:Model}
\begin{figure}
    \includegraphics[width=.48\textwidth]{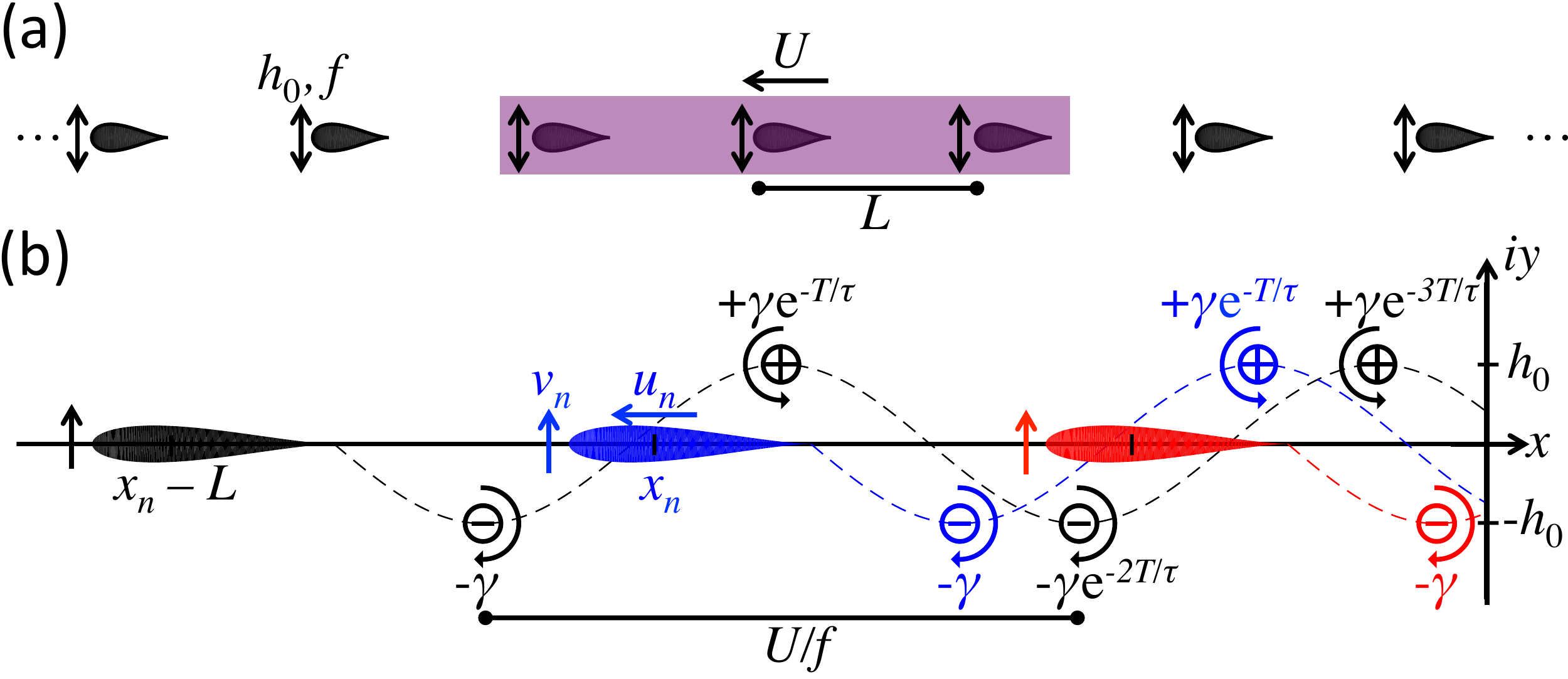}
      \caption{{\bf In-line formation of flapping swimmers.} (a) The swimmers oscillate periodically with flapping frequency $f$ and amplitude $h_0$. The distance $L$ between swimmers is fixed, and the formation's speed $U$ is determined by the balance of drag and thrust on the swimmers. The swimmers interact with their nearest neighbors, as indicated by the shaded box. (b) Diagram of the theoretical model in Eq.~\eqref{Modelv0}, presented in the complex plane $z=x+iy$. The formation's dynamics is determined by the swimmer at the center, and we track its position $x_n$, horizontal velocity $u_n$ and vertical velocity $v_n = (-1)^n2\pi fh_0$ at the midplane $y=0$. Vortices of positive (negative) circulation $\pm \gamma$ are shed from the swimmers' trailing edges at the peaks (troughs) $y=\pm h_0$ of their trajectories, indicated by the dashed curves.}
      \label{Fig:schematic}
\end{figure}

We first consider the simplest model for a school: an infinite line of swimmers modeled as heaving rigid wings driven periodically with prescribed vertical position $y(t)=h_0\sin(2\pi ft)$, and coupled by nearest-neighbor interactions, as shown in Fig.~\ref{Fig:schematic}a. We assume the swimmers to be separated by a fixed distance $L$, so the only degree of freedom is the formation's speed $U$. Our goal is to construct the governing equations for this system, and determine the dependence of $U$ on the kinematic parameters $h_0$ and $f$.

The Reynolds number of the flow around the wings used in experiments~\cite{Becker} is typically large, $\text{Re}=Uc/\nu\approx 10^{2}-10^5$, where $c$ is the chord length and $\nu$ the fluid's kinematic viscosity. Such a flow is complex and difficult to quantitatively characterize, both experimentally and numerically. We thus make the simplifying assumption that the flow is two-dimensional, inviscid and incompressible, and that the flow structures may be approximated by point vortices shed from the swimmers' trailing edges at the extrema of their trajectories (Fig.~\ref{Fig:schematic}b). These vortices mediate the interactions between swimmers. We fix the vortex strength %\footnote{We employ the convention of Streitlien \& Triantafyllou~\cite{Streitlien}, wherein a vortex strength $\gamma < 0$ denotes a vortex of positive circulation $\Gamma=-2\pi\gamma$.} 
as $\gamma =(C_v/2\pi)\int_0^T\dot{y}(t)^2\,\mathrm{d}t=C_v\pi h_0^2f/2$, where $C_v$ is a free parameter~\cite{Schnipper,Buchholz}. We also assume that the vortex strength decays exponentially over a timescale $\tau$, an assumption that accounts for turbulent breakdown of vortical structures at high Reynolds number~\cite{Daghooghi,Sophie,Higdon}. Both of these assumptions are discussed in detail in Supplemental Material \S{}II B. %\ref{App:Param}. 
We are primarily concerned with trajectories for which the swimming speed is much larger than the characteristic advection speed of vortices~\cite{MTFluid}, $U \gg \gamma/\lambda$, $\lambda$ being the distance between vortices of the same sign, and thus assume the vortices to be stationary.

We now construct evolution equations for the swimmers' position and velocity, \markblue{with the relevant variables listed in Supplemental Material Table I}. Since the swimmers interact through vortices shed by their nearest neighbors, and the distance between them is fixed, the formation's trajectory is determined by that of a single swimmer. Instead of modeling the swimmer's continuous-time motion, we evolve its horizontal position $x_n$ and velocity $u_n$ on the midplane $y=0$ at the discrete times $t_n=nT$, where $T=1/(2f)$ is the flapping half-period. At each time step, the swimmer sheds a point vortex of positive (negative) circulation at the peak (trough) of its trajectory for odd (even) $n$, which generates the characteristic reverse von K\'{a}rm\'{a}n wake~\cite{Triantafyllou_Review} of a self-propelling swimmer (Fig.~\ref{Fig:schematic}b, Supplemental Movie 1). The swimmer moves under the influence of two forces: a drag $F_D(u)$, and a propulsive thrust $F_x[x_n,u_n,v_n,\omega_n(z)]$, where $\omega_n(z)$ is the fluid vorticity in the complex plane $z=x+iy$ and $v_n=(-1)^n2\pi fh_0$ is the swimmer's vertical velocity. The equations of motion are thus
\begin{align}
&u_{n+1}=u_n+\frac{T}{m_{\text{e}}}\left(F_D(u_n)+F_x\left[x_n,u_n,v_n,\omega_n(z)\right]\right),\nonumber \\
&x_{n+1}=x_n+u_{n+1}T,\nonumber \\
&\omega_{n+1}(z)=\omega_n(z)\mathrm{e}^{-T/\tau}+(-1)^n\gamma\sum_{j=-1}^1\delta\left(z-(z_{n+1}+jL)\right),\nonumber \\
&\text{where }z_{n+1}=2a+\hat{x}_{n+1}+i(-1)^nh_0,\label{Modelv0}
\end{align}
the trailing edge of a swimmer centered at the origin is at $z=2a$, $\hat{x}_{n+1} = (x_{n+1}+x_n)/2$ and $m_{\text{e}}$ is the swimmer's effective mass per unit span, defined in Supplemental Material \S{}II. %\ref{App:NDim}. 
We impose the boundary-layer drag law $F_D(u)=-C_D\rho\sqrt{c\nu}|u|^{1/2}u$, where $\rho$ is the fluid density and $C_D$ the drag coefficient. % Check this against Silas PoF 2012 formula?
Crucially, the propulsive force $F_x$ depends on the swimmers' dynamically generated vorticity field $\omega_n(z)$.

We compute the propulsive force $F_x$ using the method detailed in Supplemental Material \S{}I. %\ref{App:FT}. %, we employ the approach of Streitlien \& Triantafyllou~\cite{Streitlien}
In summary, we model the swimmers as up-down symmetric wings in the complex plane. Such a wing is represented through the action of the so-called Joukowski map $J(\zeta) = \zeta + \zeta_c  + a^2/(\zeta + \zeta_c)$ on a circle of radius $r = |a-\zeta_c|$ in the $\zeta$-plane, where $\zeta_c\in\mathbb{R}$ sets the swimmers' vertical thickness $d_{\text{f}} = \max_{|\zeta|=r}[\text{Im}(J(\zeta))]$ and $a\approx c/4$ roughly sets the chord length. The central vortex strength $\gamma_c$ is obtained by imposing the Kutta condition at the swimmer's sharp trailing edge $z=2a$, which ensures that the flow remains smooth there. To make the calculation tractable, we assume that the swimmers influence each other only through their shed vortices and not through their body dynamics, an assumption we expect to be valid in the regime $L \gg c$. We thus obtain an explicit form for the iterated map~\eqref{Modelv0} that evolves the swimmer's position and velocity, given in Supplemental Material Eq. (12). %~\eqref{ModelOp3}. 

\markblue{We note that a similar method was used by Ramananarivo {\it et al.}~\cite{Sophie} to calculate the hydrodynamic force on a wing due to shed point vortices. However, that work neglected consideration of the wings' dynamics, which is explicitly accounted for by our iterated map~\eqref{Modelv0}. For this reason, our work goes beyond theirs in two significant ways. First, we are able to assess the stability of steady schooling states, which yields important information about which states are observed experimentally~\cite{Becker} and thus which lattice configurations are hydrodynamically optimal. Second, we are able to capture time-dependent schooling states, which we find to emerge naturally in 1D flocks (\S\ref{Sec:1DStab}). Moreover, while the prior work was limited to a 1D configuration, modeling 2D diamond lattices requires consideration of time-dependent schooling states, as shown in \S\ref{Sec:Diamond}.}

\markblue{Our theoretical model for the flow field generated by a flapping wing exhibits satisfactory agreement with Becker {\it et al.}'s measurements~\cite{Becker} using particle image velocimetry, with upstrokes and downstrokes producing upward and downward fluid flows (Supplemental Fig. 1). %~\ref{Fig:flow}
Our model thus exhibits qualitative similarity with their {\it ad hoc} 1D kinematic model (i.e. swimmers' inertia is neglected and forces directly determine the swimming speed), which posits an interaction force between swimmers that oscillates sinusoidally on the flapping period and also decays in time. However, \markred{our model} is derived from a detailed physical description of vortex-induced fluid forces, thus permitting quantitative comparison to experimental data (\S\ref{Sec:1DStab}) and treatment of 2D formations (\S\ref{Sec:Speedup}). We also explicitly incorporate the swimmers' inertia, allowing for an assessment of the stability of steady schooling states (\S\ref{Sec:1DStab}).}

Our considerable simplification of the flow structures allows the formation's dynamics to be analyzed mathematically. Moreover, simulation of the governing equations is inexpensive, which allows us to assess the dependence of the dynamics on the kinematic parameters $h_0$ and $f$. As shown in Supplemental Material \S{}II, %\ref{App:NDim}, 
the governing equations can be written in the dimensionless form
\begin{align}
&x_{n+1}=x_n+u_{n+1},\nonumber \\
&u_{n+1}=u_n+F_D(u_n)-F_0\nonumber \\
&\phantom{=}-\sum_{j=-1}^1\sum_{k=-\infty}^nG\left(x_n-\hat{x}_k-jL,u_n,n-k\right),\label{ModelSimp}
\end{align}
where $G$ is specified in Supplemental Material Eq. (17). %~\eqref{Gdef}. 

\section{Comparison with experiment}\label{Sec:1DStab}

We seek steadily translating solutions to Eq.~\eqref{ModelSimp} and assess their stability, first with the goal of rationalizing the experimental observations of Becker {\it et al.}~\cite{Becker}. The analysis of Eq.~\eqref{ModelSimp} is nontrivial due to the temporal nonlocality of the formation's dynamics: updating the velocity $u_n$ requires knowledge of the swimmer's history, which is a generic feature of flow-induced interactions at high Reynolds number. 

Substituting the steady state $x_n=Un,u_n=U$ into Eq.~\eqref{ModelSimp}, we find that $U$ satisfies the algebraic equation
\begin{align}
F_D(U)=F_0+\sum_{j=-1}^1\sum_{m=0}^{\infty}G(U(m+1/2)-jL,U,m).\label{UEqn}
\end{align}
This equation is solved numerically using a bisection method. The linear stability analysis of such steady state solutions is given in Supplemental Material \S{}III. %\ref{App:Stab}. 
In summary, we linearize Eq.~\eqref{ModelSimp} around the steady state solution, and find the eigenvalues of the linear stability problem using the discrete Laplace transform. We show that the eigenvalues are given by the zeros of the function
\begin{align}
\mathcal{F}(z)&= (z-1)\left(z-\left(1+F_D^{\prime}(U)-g_u\right)\right)-\frac{1}{2}\mathcal{G}(z)(z+1)\nonumber \\
&\phantom{=}+g_xz,\label{Fdef}
\end{align}
where the constants $g_x,g_u$ and function $\mathcal{G}(z)$ are specified in Supplemental Material \S{}III. %\ref{App:Stab}.
The steady state solution is stable if all of the roots of $\mathcal{F}(z)$ lie inside the unit disc in the complex plane, $|z|<1$, and is unstable otherwise. We use a numerical contour integration method~\cite{Delves_1966} to find the roots of $\mathcal{F}(z)$ in the annular region $1 < z < R$, where $R$ is a sufficiently large number. The stability properties of the steady state solution $u_n=U$ are dictated by the location of the root $z^*$ of $\mathcal{F}(z)$ that is largest in magnitude.

\begin{figure*}%* for full width
\centering
     \includegraphics[width=1\textwidth]{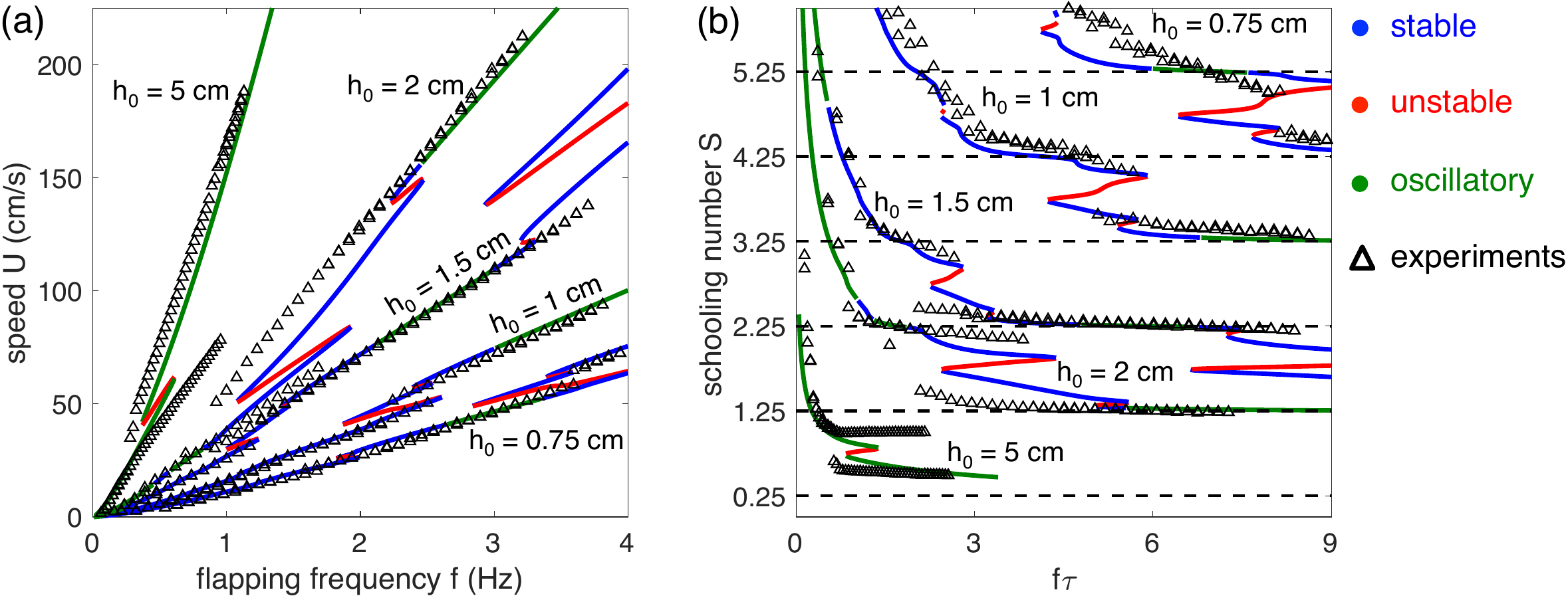}
      \caption{{\bf Model validation.} Comparison between the experimental data of Becker {\it et al.}~\cite{Becker} (triangles) and the theoretical predictions of Eq.~\eqref{ModelSimp}--\eqref{UEqn} (curves). In panel (a), the formation's speed $U$ is plotted as a function of the flapping frequency $f$. In panel (b), the speed $U$ is plotted in terms of the schooling number $S = Lf/|U|$, and the flapping frequency is made nondimensional by the vortex decay time $\tau$. The theoretically predicted solutions are color-coded according to their stability, as outlined in \S\ref{Sec:1DStab}: blue indicates stable solutions, red unstable solutions, and green oscillatory solutions.}
      \label{Fig:data}
\end{figure*}

Figure~\ref{Fig:data} shows the comparison between theory (curves) and experiment~\cite{Becker} (triangles) for an in-line formation. In the experiments of Becker {\it et al.}~\cite{Becker}, the distance $L$ between swimmers is fixed, while the flapping frequency $f$ and amplitude $h_0$ are varied. The measured swimming speed $|U|$ is shown in Fig.~\ref{Fig:data}a. The data in Fig.~\ref{Fig:data}b are plotted in terms of the schooling number $S = Lf/|U|$, which denotes the number of wavelengths separating the swimmers~\cite{Becker}. That is, integer values of $S$ indicate trajectories for which the swimmers traverse identical paths, and half-integer values indicate trajectories for which neighboring swimmers traverse paths that are perfectly out-of-phase. The steady state solutions predicted by Eq.~\eqref{UEqn} are color-coded according to their stability: specifically, blue denotes stable states; red denotes unstable states for which $\text{Re}(z^*)>0$ and $\text{Im}(z^*)=0$; and green denotes oscillatory states, which may destabilize via either a flip bifurcation ($\text{Re}(z^*)<0$ and $\text{Im}(z^*)=0$) or a Neimark-Sacker bifurcation ($\text{Im}(z^*)\neq 0$). %cyan denotes unstable states that destabilize via a flip bifurcation, $\text{Re}(z^*)<0$ and $\text{Im}(z^*)=0$; and green denotes unstable states that destabilize via a Neimark-Sacker bifurcation, $\text{Im}(z^*)\neq 0$. 
\markblue{The physical significance of these instabilities is explained at the end of this section.}

%* for full width

The theoretical predictions in Fig.~\ref{Fig:data} exhibit generally excellent agreement with the experimental data of Becker {\it et al.}~\cite{Becker}. At the lowest flapping amplitude considered, $h_0 = 0.75$ cm, the agreement between theory and experiment is less good, presumably owing to the breakdown of the point-vortex approximation at low flapping amplitudes. We note that three free parameters, namely, $C_D$ (drag coefficient), $\tau$ (vortex time decay) and $C_v$ (initial vortex strength), are chosen once to best fit all of the experimental data across flapping frequency $f$ and amplitude $h_0$. \markblue{The numerical values of these parameters exhibit good agreement with results in the existing literature, as detailed in Supplemental Material \S{}II B. %\ref{App:Param}. 
Indeed, our fit value $\tau\approx 2$ seconds compares well with the value $\tau\approx 5$ seconds inferred from the data of Newbolt {\it et al.} (\cite{Newbolt}, Supporting Information), who measured the temporal decay of the flow generated by a flapping wing in a water tank.}  %Similarly, the disagreement between theory and experiment at the largest flapping amplitude considered, $h_0 = 5$ cm, may be ascribed to the curvature of the experimental setup, as the high-speed swimmers traverse the cylindrical tank in as little as one or two flaps. 
Generally, the correspondence between theory and experiment makes clear that our simplified approach to modeling the flow and the swimmers' dynamics captures the key features of the hydrodynamic interactions between swimmers at high Reynolds number.

The experiments of Becker {\it et al.}~\cite{Becker} provided evidence for the bistability of steady states, and the emergence of coexisting \lq\lq slow modes\rq\rq\, and \lq\lq fast modes\rq\rq\, for the same flapping frequency $f$ and amplitude $h_0$. Our theoretical predictions explain these observations in terms of the stability properties of the steady state solutions. Indeed, branches of stable (blue) solutions are separated by unstable (red) branches, as shown in Fig.~\ref{Fig:data}. Specifically, schooling numbers $S\approx s + 1/4$ for $s\in\mathbb{N}$ are favored by the system when the wings are moving fast due to their large flapping frequency, a regime in which hydrodynamic interactions are the strongest. The unstable branches typically run through schooling numbers $S\approx s + 3/4$, which are thus avoided by the system. The emergence of these schooling numbers is explained analytically in Supplemental Material \S{}II A. %\ref{App:FastAsym}. 
We note that the oscillatory structure of the predicted solutions is a consequence of the hydrodynamic coupling between swimmers, as the speed is a monotonically increasing function of the flapping frequency $f$ for an isolated swimmer. %Taken together, the experimental and theoretical results in Fig.~\ref{Fig:data} indicate that swimmer in a school may benefit from collective hydrodynamic interactions, as they can experience speedup when together.

In addition to explaining the bistability of steady states, the linear stability analysis in Fig.~\ref{Fig:data} hints that in-line formations may show unsteady behavior, which is confirmed by numerical simulations of Eq.~\eqref{ModelSimp}. Specifically, Supplemental Fig. 2a %\ref{Fig:dynamics}a 
shows a simulation conducted in a parameter regime in which the steady state solution $u_n = U$ goes unstable via a flip bifurcation, so the swimmer's velocity oscillates on the flapping period. In experiments, one would thus expect the swimming speed to change appreciably during a single flap, but to be roughly the same at the start of each flap.  More interestingly, Supplemental Fig. 2c %\ref{Fig:dynamics}c 
shows that a steady schooling state may also undergo a Neimark-Sacker bifurcation, in which the swimming speed oscillates over a period long relative to the flapping period, $T_{\text{osc}}\approx\pi\left[\text{Im}\left(\log z^*\right)\right]^{-1}\approx 6T$ for the parameter regime explored here. Due to its simplicity, our model furnishes testable predictions for the manner in which steady schooling states destabilize, and the parameter regimes in which they do so. 

\section{Comparative analysis of different formations}\label{Sec:Speedup}

Having benchmarked our theoretical model against experimental data, we analyze how hydrodynamic interactions impact the performance of different lattice formations. %now investigate how the lattice geometry influences the hydrodynamic advantage gained by swimmers in a school. 
We consider two performance measures: the formation's speed $U$ and cost of transport $C$, and compare these with the corresponding speed $U_0$ and cost of transport $C_0$ of a single isolated swimmer. The cost of transport $C = \langle P_n/|u_n|\rangle$ is a \lq\lq gallons-per-mile\rq\rq\, measure of efficiency that quantifies the formation's energy consumption per unit distance~\cite{Tokic_COT}, where $P_n$ is the instantaneous \markblue{mechanical power output of} the swimmer at time $t_n = nT$ and $\langle\cdot\rangle$ denotes a time average. A formula for $P_n$ is given in Supplemental Material \S{}IV. %\ref{App:COT}.

We compute the speed $U$ and cost of transport $C$ of in-line (\S\ref{Sec:Line}), phalanx (\S\ref{Sec:Phalanx}), rectangular lattice (\S\ref{Sec:Rect}), and diamond lattice (\S\ref{Sec:Diamond}) formations as a function of the distance between swimmers. Our goal is to identify the lattice formations that maximize the speed and minimize the cost of transport relative to that of a single swimmer, values $U/U_0 > 1$ and $C/C_0 < 1$ indicating a benefit due to collective hydrodynamic interactions. In this section, we restrict our attention to a single representative set of flapping kinematics, $f = 1.5$ Hz and $h_0 = 1.5$ cm, for which $h_0/c= 0.25$ and $\text{St} \equiv 2h_0 f/U_0\approx 0.1$, the low Strouhal number regime $\text{St}\ll 1$ being biologically relevant for fish schools~\cite{Triantafyllou_Review}. All distances are reported in units of the swimmer's approximate body length $4a$. 

%Following Weihs~\cite{Weihs1,Weihs2}, our goal is to determine the extent to which swimmers benefit from hydrodynamic interactions in a 2D geometry. 
 
\subsection{In-line formation}\label{Sec:Line}

\begin{figure}
    \includegraphics[width=0.48\textwidth]{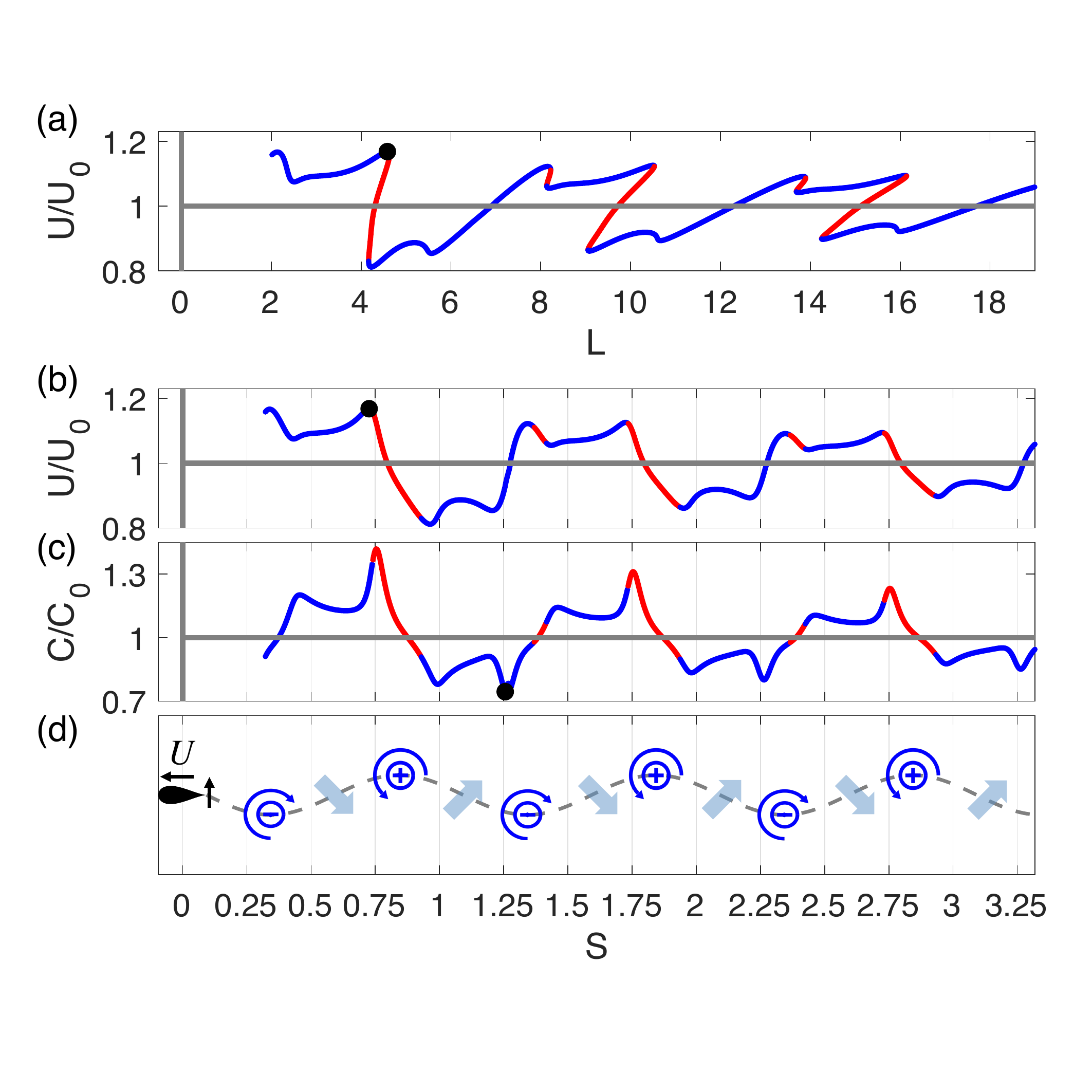}
      \caption{{\bf Speedup and cost of transport in an in-line formation.} (a) Speed $U$ of an in-line formation of swimmers separated by a distance $L$, as predicted by Eq.~\eqref{UEqn}, compared to the speed $U_0$ of a single isolated swimmer. The fastest formation has $U/U_0 = 1.17$ for $L = 4.6$, corresponding to $S = 0.73$ (black dot). (b) Speedup $U/U_0$ from panel (a) plotted in terms of the schooling number $S$. (c) Cost of transport $C$ compared to that of a single swimmer, $C_0$. The most efficient state has $C/C_0 = 0.75$ for $L = 6.3$, corresponding to $S = 1.26$ (black dot). In panels (a)--(c), blue denotes stable or oscillatory states, red denotes unstable states and the black dots indicate the fastest (panels (a) and (b)) and most efficient (panel (c)) states. (d) Fluid flow (thick blue arrows) associated with the reverse von K\'{a}rm\'{a}n street shed by a flapping swimmer. For a swimmer (not shown) located at a dimensionless distance $S$ downstream, regions of downflow (upflow) are typically associated with speedup (slowdown) and higher (lower) cost of transport in panels (b) and (c). Dashed line indicates the upstream swimmer's trajectory. }
      \label{Fig:LineSpeed}
\end{figure}

We first solve Eq.~\eqref{UEqn} to find the dependence of the swimming speed $U$ on the streamwise distance $L$ between swimmers in a line. As in \S\ref{Sec:1DStab}, we assume the formation's dynamics to be dominated by nearest-neighbor interactions in the streamwise direction. The results are shown in Fig.~\ref{Fig:LineSpeed}. As expected from the discussion in \S\ref{Sec:1DStab}, a slow mode ($U/U_0 < 1$) and fast mode ($U/U_0  > 1$) may coexist for a given value of $L$ (Fig.~\ref{Fig:LineSpeed}a). The maximum speedup of 17\% corresponds to a state with $S=0.73$, % S = 0.726, occurs at $L = 4.585$, 
while the largest slowdown by 19\% corresponds to a state with $S=0.96$ (Fig.~\ref{Fig:LineSpeed}b). %L = 4.266, S = 0.963 
The lowest cost of transport is $C/C_0 = 0.75$ and %0.74517 for P2 state
corresponds to a state with $S = 1.26$, %S = 1.2559, and occurs at $L = 6.311
indicating a maximum energy savings of 25\%, while the highest cost of transport $C/C_0 = 1.42$ corresponds to a state with $S=0.75$ (Fig.~\ref{Fig:LineSpeed}c). %S = 0.754 $L = 4.5804
Plotting $U/U_0$ as a function of $S$ (Fig.~\ref{Fig:LineSpeed}b) shows that states with $S\gtrsim s+1/4$ and $S\lesssim s+3/4$ typically have the highest speeds, whereas those with $S\lesssim s$ and $S\lesssim s+1/4$ have the lowest. Conversely, Fig.~\ref{Fig:LineSpeed}c shows that states with $S\lesssim s+3/4$ typically have the highest cost of transport, whereas those with $S\lesssim s$ and $S\lesssim s+1/4$ have the lowest. Comparing Fig.~\ref{Fig:LineSpeed}b and~\ref{Fig:LineSpeed}c, we observe that high-speed states ($U/U_0 > 1$) are typically associated with an increased cost of transport ($C/C_0 > 1$), indicating a tradeoff between speed and energy consumption.

To understand the oscillatory dependence of $U/U_0$ on $L$, we derive an approximate form for the thrust $F^{\text{v}}_x$ and lift $F_y^{\text{v}}$ on a swimmer due to the vortices shed by its neighbors, assuming that the vortices (with positions $z_k\in\mathbb{C}$ and strengths $\gamma_k$) are far from the body, $|z_k| \gg r$. In Supplemental Material \S{}I, %\ref{App:FT}, 
we show that
\begin{align}
F^{\text{v}}_x &\approx 4\pi\rho r VV_{\text{f}}\quad\text{and}\quad F^{\text{v}}_y\approx -4\pi\rho r UV_{\text{f}},\nonumber \\
\text{where}\quad V_{\text{f}}&=\text{Im}\left.\left(\sum_k\frac{i\gamma_k}{z-z_k}\right)\right|_{z=0}\label{FApprox}
\end{align}
is the vertical velocity of the flow induced by the neighboring swimmers' shed vortices and $(U,V)$ is the swimmer's velocity.
%That is, the horizontal force on a swimmer due to its neighbors is roughly proportional to the product of its own vertical velocity $V$ and the vertical velocity $V_{\text{f}}$ of the flow induced by its neighbors' shed vortices. 
Figure~\ref{Fig:LineSpeed}d shows a schematic of the reverse von K\'{a}rm\'{a}n (thrust) wake and associated fluid flow generated by a flapping swimmer. %, with the spatial coordinate expressed in terms of the schooling number $S$. 
A swimmer moving upward ($V>0$) in its neighbor's wake would experience a constructive interaction force $F^{\text{v}}_x  < 0$ and thus a speedup ($U/U_0 > 1$) for $s+1/4\lesssim S\lesssim s+3/4$, and the opposite for $s+3/4\lesssim S \lesssim s+ 5/4$, according to Eq.~\eqref{FApprox}. We note that these conclusions may be justified mathematically by adapting the argument in Supplemental Material \S{}II A, %\ref{App:FastAsym}, 
in which we derive an approximate form for the interaction force~\eqref{FApprox} in the biologically-relevant low Strouhal number limit $\text{St}\ll 1$~\cite{Triantafyllou_Review}.
%, as was done in prior work~\cite{Sophie}. This simple physical picture offers a qualitative explanation for the the oscillatory dependences of the data in Fig.~\ref{Fig:LineSpeed}a--b). 

A similar argument  may be used to qualitatively understand the tradeoff between speed and cost of transport. A wing moving up ($V>0$) in  a high-speed state ($U/U_0 > 1$) experiences a downward flow from its neighbor's wake ($V_{\text{f}}<0$), which implies $F_y^{\text{v}}<0$ by Eq.~\eqref{FApprox}. High-speed states are thus typically associated with an increased power consumption, as the wing's vertical motion is opposed by its neighbor's induced flow. In the low Strouhal number limit $\text{St} \ll 1$, the increase in cost of transport $C$ due to the increased power consumption dominates the decrease in $C$ associated with the higher speed, as shown in Supplemental Material \S{}IV A. %\ref{App:COT1D}. 

\markblue{We may use the foregoing arguments to draw conclusions about an in-line formation in which the nearest neighbors flap perfectly out-of-phase with respect to each other. For such a configuration, the right-hand sides of the formulae in Eq.~\eqref{FApprox} for $F_x^{\text{v}}$ and $F_y^{\text{v}}$ would simply have their signs reversed. We thus expect a speedup and larger cost of transport for schooling numbers $s+3/4\lesssim S\lesssim s+5/4$, and a slowdown with lower cost of transport for $s+1/4\lesssim S\lesssim s+3/4$. That is, Figs.~\ref{Fig:LineSpeed}b and~\ref{Fig:LineSpeed}c would be qualitatively unchanged, apart from a shift of the horizontal axis, $S\rightarrow S+1/2$.}

\subsection{Phalanx}\label{Sec:Phalanx}
Motivated by the experimental observations of Ashraf {\it et al.}~\cite{AshrafPNAS}, we now consider a phalanx of swimmers: infinitely many swimmers equally spaced by a distance $d$ in the lateral ($y$) direction and, following Weihs~\cite{Weihs2}, flapping in antiphase with respect to their neighbors, as shown in Fig.~\ref{Fig:Phalanx2} and Supplemental Movie 2. As discussed in Supplemental Material \S{}V A, %\ref{App:Antiphase}, 
Weihs~\cite{Weihs2} and St\"{o}cker~\cite{Stocker} argued that in-phase flapping would result in an increased drag force on the downstream fish, due to the anomalously large induced velocity in the $y$-direction. Such an argument is in agreement with the experimental observations of Ashraf {\it et al.}~\cite{AshrafINT} who found that pairs and triplets of red nose tetra fish preferentially flap in antiphase with respect to their lateral neighbors. We also restrict our attention to the parameter regime $d> d_{\text{min}}\equiv 2(h_0 + d_{\text{f}})$ %$d_{\text{f}}  < h_0 < d/2 - d_{\text{f}}, $t = \max \text{Im}[z_{\text{foil}}] = 0.0772$
to ensure that the swimmers do not collide with each other during a flapping cycle. 

A straightforward extension of the iterated map presented in \S\ref{Sec:Model} permits consideration of this formation, as detailed in Supplemental Material \S{}V. %\ref{App:2D}. 
Following the procedure detailed in \S\ref{Sec:1DStab}, we find that the swimming speed $U$ satisfies an algebraic equation of the form~\eqref{UEqn} with $L = 0$. The interaction function $G$ describes the hydrodynamic thrust due to a % column
side-by-side arrangement of swimmers flapping in antiphase, and is defined in Supplemental Material Eq. (41). %~\eqref{GdefCol2}. 
The solutions to this equation are shown in Fig.~\ref{Fig:Phalanx2}. The phalanx formation evidently does not exhibit the multi-stability of steady states seen for in-line formations (\S\ref{Sec:Line}). By generalizing the linear stability analysis of steady state solutions presented in \S\ref{Sec:1DStab}, we find that the steady state is stable for all values of $d$. As shown in Fig.~\ref{Fig:Phalanx2}, the formation exhibits a slight speedup for all values of $d$, with the maximum speedup of roughly 5\% occurring when the swimmers are most tightly packed, $d = d_{\text{min}}$. However, such a formation also increases the cost of transport by roughly 4\%, with the cost of transport decreasing as $d\rightarrow\infty$. Similar to the 1D formations discussed in \S\ref{Sec:Line}, the phalanx formation exhibits a tradeoff between speed and cost of transport, although both measures exhibit relatively small variations over a range of values $d$.

\begin{figure}
    \includegraphics[width=0.48\textwidth]{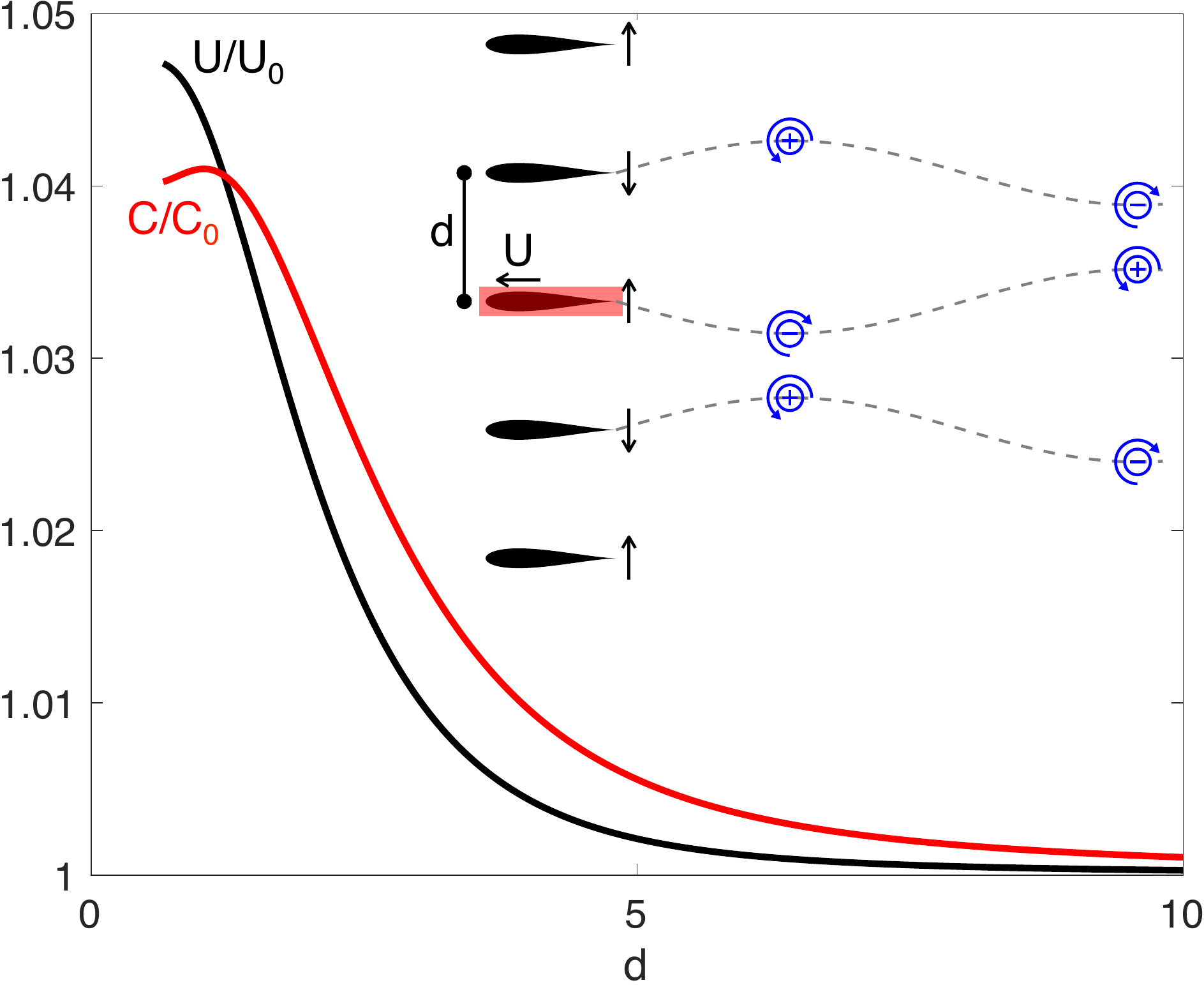}%{WingVortex_Phalanx_ExactSoln1_P1-f1p5h1p5n100_with_COT}
      \caption{{\bf Speedup and cost of transport in a phalanx.} The phalanx formation consists of infinitely many swimmers equally spaced by a distance $d$ in the lateral direction, flapping in antiphase with respect to their neighbors. Black arrows indicate the swimmers' instantaneous vertical velocity. Dashed lines indicate the swimmers' trajectories, with associated shed vortices (blue). The speed (black curve) and cost of transport (red curve) evidently decrease as the lateral distance $d$ between swimmers is increased. }
      \label{Fig:Phalanx2}
\end{figure}

\subsection{Rectangular lattice}\label{Sec:Rect}
\begin{figure}
    \includegraphics[width=0.48\textwidth]{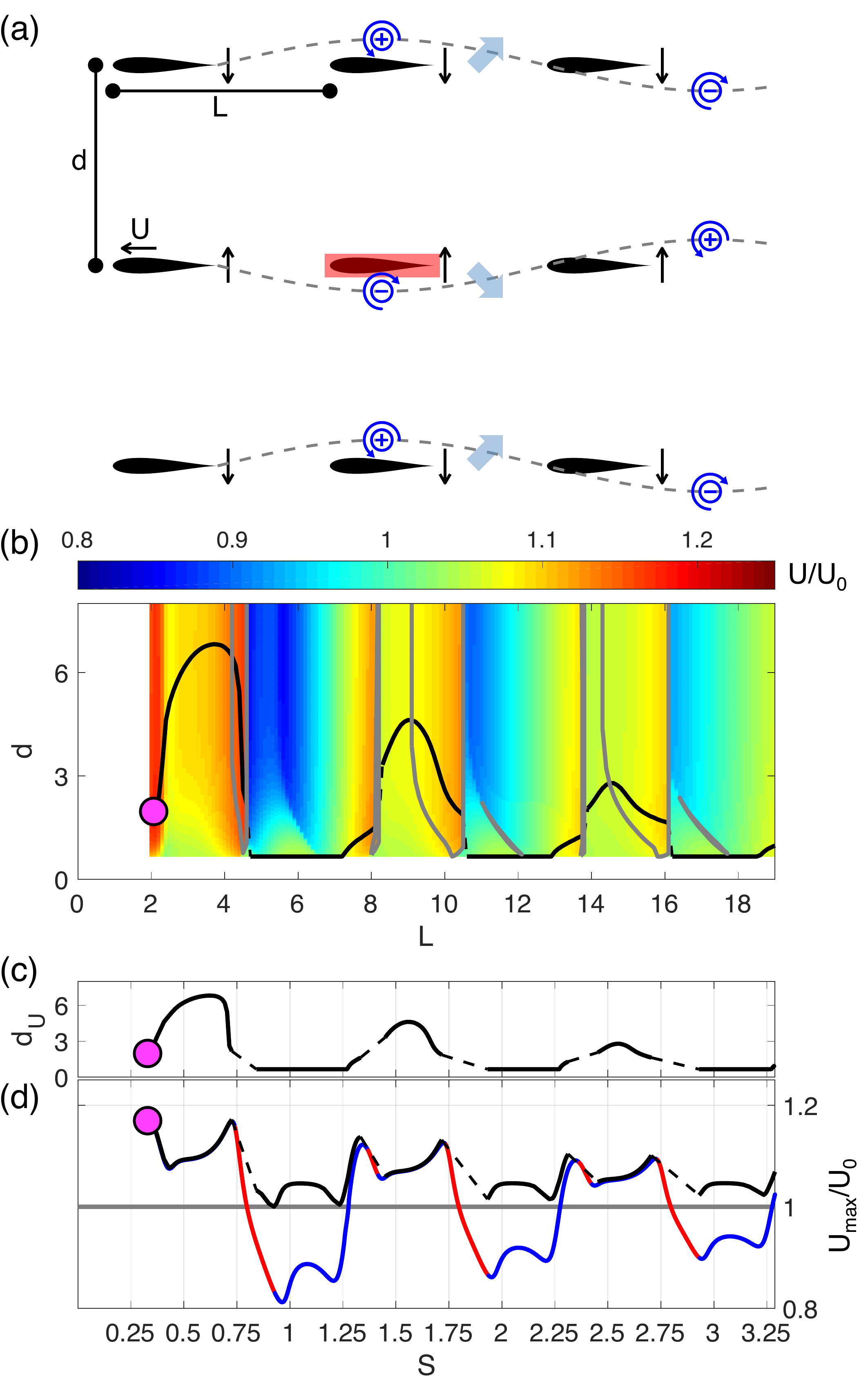}%{RectFig5.pdf}
      \caption{{\bf Speedup in a rectangular lattice.} (a) Schematic of the rectangular lattice of swimmers that maximizes the speedup $U_{\text{av}}/U_0$. This state has $U_{\text{av}}/U_0=1.18$ for $L = 2.1$ and $d = 1.94$, corresponding to $S = 0.33$, and is indicated by the pink dot in panels (b)--(d). The formation's dynamics is determined by that of the swimmer at the origin (red box), and black arrows indicate the swimmers' instantaneous vertical velocity. Dashed lines indicate the trajectories of the swimmers in the upstream column, with associated vortices (blue) and fluid flows (thick blue arrows). (b) The colormap shows the normalized time-averaged speed $U_{\text{av}}/U_0$ as a function of $L$ and $d$, based on numerical simulations of rectangular lattice formations. The black curve $d_U(L)$ indicates the optimal lateral spacing as a function of the streamwise spacing $L$. Multiple coexisting states, obtained by changing the initial conditions, are found in the regions bounded by the gray curves. (c) Optimal lateral spacing $d_U(L)$ plotted as a function of the schooling number $S$. (d) The associated speedup $U_{\text{max}}/U_0$ (black), superimposed on top of the results for an in-line formation from Fig.~\ref{Fig:LineSpeed}b. The dashed portions of the black curves in panels (b)--(d) are guides to the eye.}
      \label{Fig:Rect}
\end{figure}

We now consider the rectangular lattice of swimmers shown in Fig.~\ref{Fig:Rect}a and Supplemental Movie 3: swimmers separated by a streamwise distance $L$ and a vertical distance $d$, starting at the positions $z = jL+ikd$ for $j,k\in\mathbb{Z}$. Swimmers %in a row (column)
flap in phase (antiphase) with respect to their streamwise (lateral) neighbors. As in \S\ref{Sec:Model}, a swimmer at the origin (red box in Fig.~\ref{Fig:Rect}a) interacts with the swimmers in the neighboring columns at $x=0,\pm L$. Note that the lattice is effectively an in-line formation in the limit $d\rightarrow\infty$. In Supplemental Material \S{}V, %\ref{App:2D}, 
we show that the steady speed $U$ satisfies the algebraic equation~\eqref{UEqn}, with the function $G$ defined in Supplemental Material Eq. (41). %~\eqref{GdefCol2}. 

We numerically  solve this equation to find the dependence of the steady speed $U$ on the geometric parameters $L$ and $d$. We then perform numerical simulations of the evolution equations, with initial conditions determined by these steady-state solutions, and compute the trajectory's time-averaged velocity $U_{\text{av}}$. Figure~\ref{Fig:Rect}b shows the normalized velocity $U_{\text{av}}/U_0$ as a function of $L$ and $d$. As with the in-line formation, there may be multiple steady-state solutions for a given set of parameters; in such cases, we perform multiple simulations %with initial conditions determined by each solution, 
and plot the time-averaged speed of the fastest state. The simulations reveal the existence of multiple coexisting states within the regions of geometric parameter space bounded by the gray curves in Fig.~\ref{Fig:Rect}b. As the lateral distance $d$ between swimmers is decreased progressively, these regions of multi-stability typically shrink, but new regions of multi-stability may also emerge.  

\begin{figure}
  %\centering
    \includegraphics[width=.48\textwidth]{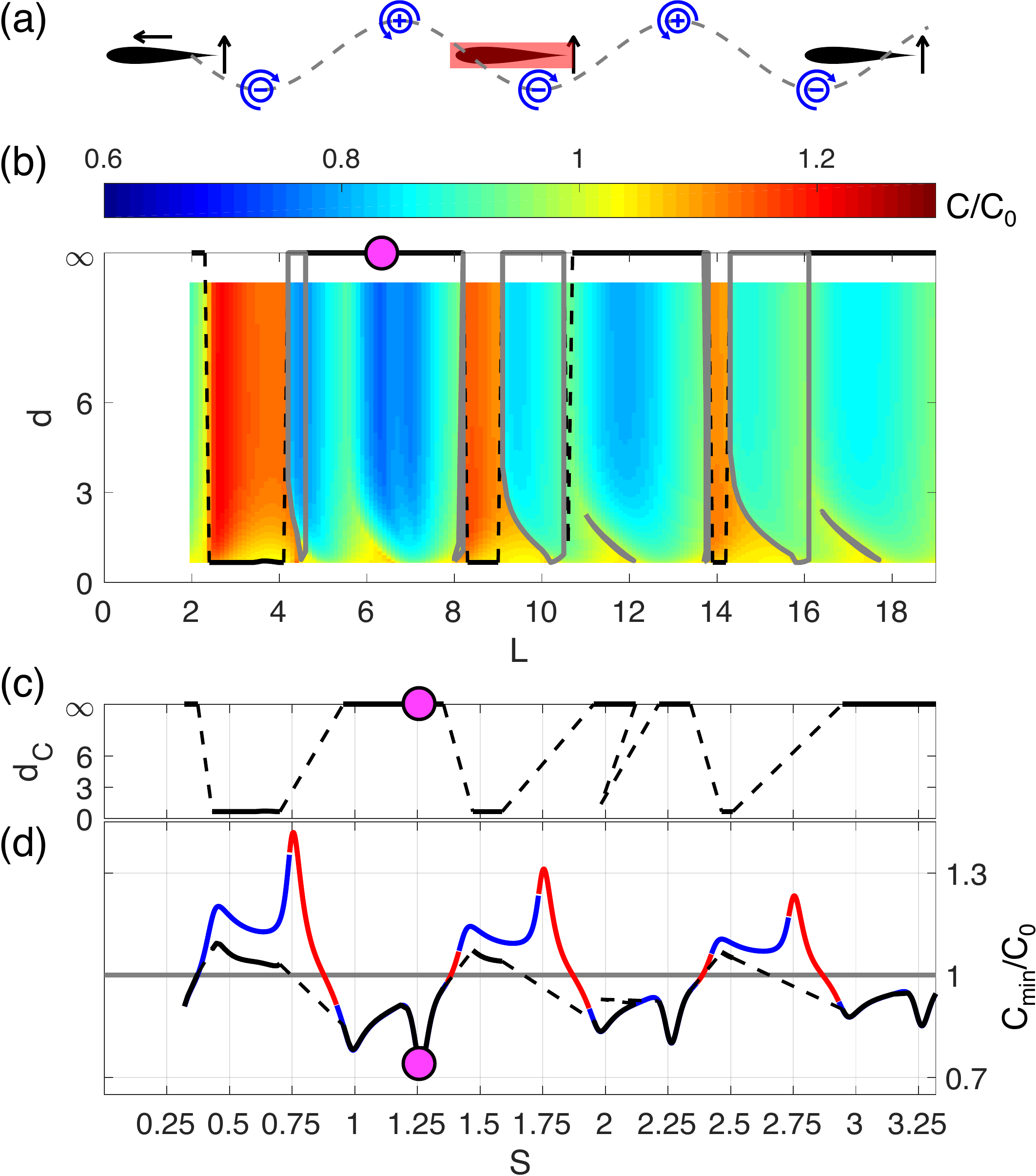}%{RectFigCOT-1.pdf}
      \caption{{\bf Cost of transport in a rectangular lattice.} (a) Schematic of the rectangular lattice of swimmers that minimizes the cost of transport $C/C_0$. The minimal cost of transport $C/C_0 = 0.75$ at $L = 6.3$ and $S = 1.26$ corresponds to that of an in-line formation (Fig.~\ref{Fig:LineSpeed}c), for which $d\rightarrow\infty$. This state is indicated by the pink dot in panels (b)--(d). (b) The colormap shows the normalized cost of transport $C/C_0$ as a function of $L$ and $d$, based on numerical simulations of the rectangular lattice formation. The black curve $d_C(L)$ indicates the optimal lateral spacing as a function of the streamwise spacing $L$.   (c) Optimal lateral spacing $d_C(L)$ plotted as a function of the schooling number $S$. (d) The associated cost of transport $C_{\text{min}}/C_0$ (black), superimposed on top of the results for an in-line formation from Fig.~\ref{Fig:LineSpeed}c.}
      \label{Fig:Rect_COT}
\end{figure}

We find that the formation experiences a maximum speedup of 18\% for a roughly square geometry, $L=2.1$ ($S = 0.33$) and $d = 1.94$. The black curves in Fig.~\ref{Fig:Rect} indicate $d_U(L) = \arg\max_d(U_{\text{av}}(L,d)/U_0)$, the optimal lateral spacing for a given streamwise spacing, and the corresponding speed $U_{\text{max}}(L) = |U_{\text{av}}(L,d_U(L))|$. Note that, unlike the phalanx (Fig.~\ref{Fig:Phalanx2}), the greatest speedup is not necessarily achieved by packing the swimmers tightly in the $y$-direction, so $d_U(L)$ is not identically equal to $d_{\text{min}}$. A comparison between the rectangular lattice and in-line formation is shown in Fig.~\ref{Fig:Rect}d. The most salient feature is that the slow modes ($U/U_0 < 1$) for an in-line formation all exhibit speedup in the corresponding rectangular lattice with the minimum lateral spacing, $d_U(L) = d_{\text{min}}$. However, the fast modes ($U/U_0 > 1$) for an in-line formation benefit minimally from the rectangular geometry, and the corresponding optimal lateral spacing $d_U(L)$ is often much larger than $d_{\text{min}}$. Indeed, the fastest rectangular lattice formation is only marginally faster than the fastest in-line formation.

Figure~\ref{Fig:Rect_COT} shows the corresponding results for the cost of transport of a rectangular lattice of swimmers. For the regions of parameter space in which multiple states coexist, Fig.~\ref{Fig:Rect_COT}b shows the cost of transport of the most efficient one. The cost of transport assumes the minimum value for an effectively in-line formation ($d\rightarrow\infty$), for which $C/C_0 = 0.75$ and $S = 1.26$ (Fig.~\ref{Fig:LineSpeed}c). The black curves in Fig.~\ref{Fig:Rect_COT} correspond to $d_C(L) = \arg\min_dC(L,d)$, the optimal lateral spacing for a given streamwise spacing, and the corresponding cost of transport $C_{\text{min}}(L) = C(L,d_C(L))$. Unlike $d_U(L)$, which exhibits a nontrivial dependence on the streamwise spacing $L$ (Fig.~\ref{Fig:Rect}b), $d_C(L)$ typically assumes the values $d_C\approx d_{\text{min}}$ and $d_C=\infty$ (Fig.~\ref{Fig:Rect_COT}b). Figure~\ref{Fig:Rect_COT}d shows that the states for which $d_C\approx d_{\text{min}}$ typically correspond to inefficient states ($C/C_0 > 1$) for which the rectangular lattice formation affords a slight hydrodynamic advantage over the in-line formation. Conversely, the states for which $d_C=\infty$ typically correspond to efficient states ($C/C_0 < 1$) for which the lattice geometry actually increases the cost of transport, making the in-line formation the most efficient. 

Taken together, the results in Fig.~\ref{Fig:Rect}d and Fig.~\ref{Fig:Rect_COT}d make clear the dominance of streamwise interactions between swimmers in determining both the speed and cost of transport of a rectangular lattice of swimmers. The physical picture given in \S\ref{Sec:Line} may thus be used to qualitatively understand how the lattice geometry influences both the speed and cost of transport. Recall that the in-line formation typically experiences a speedup ($U/U_0 > 1$) and a decrease in efficiency ($C/C_0 > 1$) when a swimmer at $z=0$ swims up into the downflow generated by its upstream neighbor at $z=-L$ (Fig.~\ref{Fig:LineSpeed}). For such values of $L$, corresponding to schooling number $s+1/4\lesssim S\lesssim s+3/4$, the upstream swimmers at  $z=-L\pm id$ in a rectangular lattice contribute an upflow (Fig.~\ref{Fig:Rect}a), decreasing the thrust but increasing the lift according to Eq.~\eqref{FApprox}. These effects contribute to a decrease in speed but an increase in efficiency for a rectangular lattice as compared to an in-line formation, making it advantageous to increase (decrease) the lateral spacing $d$ when optimizing for speed (efficiency). Conversely, the in-line formation is relatively efficient ($C/C_0 < 1$) but slow ($U/U_0 < 1$) for values of $L$ corresponding to $s+3/4\lesssim S\lesssim s+5/4$, the parameter regime in which the upstream swimmers at $z=-L\pm id$ in the rectangular lattice generate a downflow. This contributes to an increase in speed but a decrease in efficiency, making it advantageous to decrease (increase) the lateral spacing $d$ when optimizing for speed (efficiency).

\subsection{Diamond lattice}\label{Sec:Diamond}
\begin{figure}
  %\centering
    \includegraphics[width=.48\textwidth]{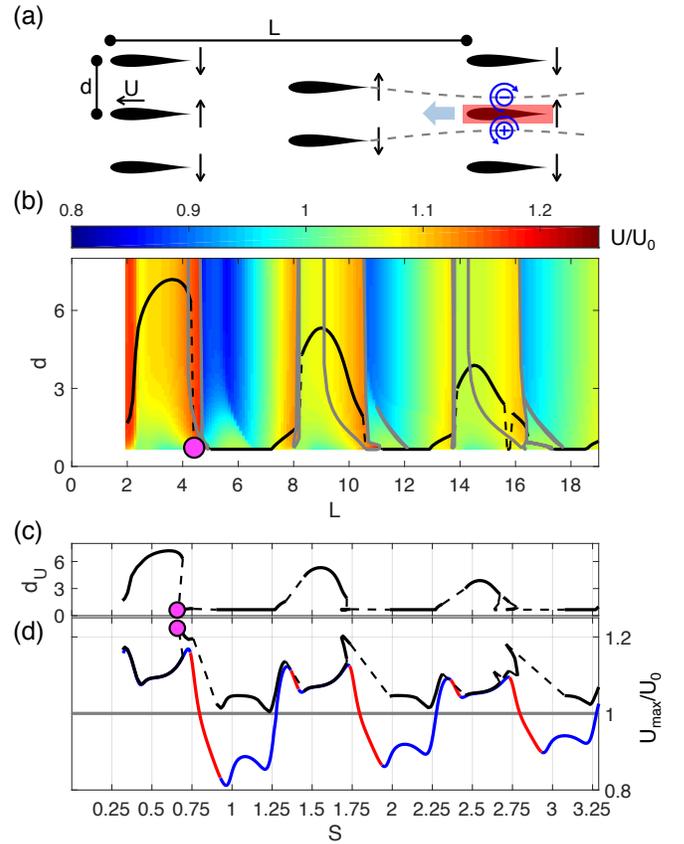}%{ArrayFig7.pdf}%{test2.pdf}
      \caption{{\bf Speedup in a diamond lattice.} The procedure described in the caption of Fig.~\ref{Fig:Rect} is repeated for a diamond lattice. The fastest state has $U_{\text{av}}/U_0=1.22$ for $L = 4.4$ and $d = d_{\text{min}}$, corresponding to $S = 0.67$. 
      %(a) Schematic of the diamond lattice of swimmers that maximizes the speedup $U_{\text{av}}/U_0$. %The lattice consists of five columns of swimmers at $x = \pm L,\pm L/2,0$; only the ones for $x\leq 0$ are shown, and forces are computed on the swimmer at the origin (red box). 
%Black arrows indicate the swimmers' instantaneous vertical velocity . Dashed lines indicate the trajectories of two upstream swimmers, with associated vortices (blue) and fluid flow (thick blue arrow). (b) Normalized time-averaged speed $U_{\text{av}}/U_0$ as a function of $L$ and $d$, based on numerical simulations of the diamond lattice. The black curve $d^*_U(L)$ indicates the optimal lateral spacing as a function of the streamwise spacing $L$. (c) Optimal lateral spacing $d^*_U(L)$ plotted as a function of the schooling number $S$. (d) The associated speedup $U_{\text{max}}/U_0$ (black), superimposed on top of the results for a linear school from Fig.~\ref{Fig:LineSpeed}b.
}
      \label{Fig:Array}
\end{figure}

We now consider the diamond lattice shown in Fig.~\ref{Fig:Array}a and Supplemental Movie 4: %columns of swimmers separated by a horizontal distance $L/2$ and a vertical distance $d$, with the columns at $x = (j+1/2)L$ shifted upward by $d/2$ for $j\in\mathbb{Z}$. 
swimmers separated by a streamwise distance $L$ and lateral distance $d$, starting at the positions $z=(j+l/2)L+i(k+l/2)d$ for $j,k\in\mathbb{Z}$ and $l\in\{0,1\}$. As in the previous section, swimmers %in a row (column) 
flap in phase (antiphase) with respect to their streamwise (lateral) neighbors, and a swimmer at the origin (red box in Fig.~\ref{Fig:Array}a) interacts with the swimmers in the neighboring columns at $x=0,\pm L/2,\pm L$.

\begin{figure}
  %\centering
    \includegraphics[width=.48\textwidth]{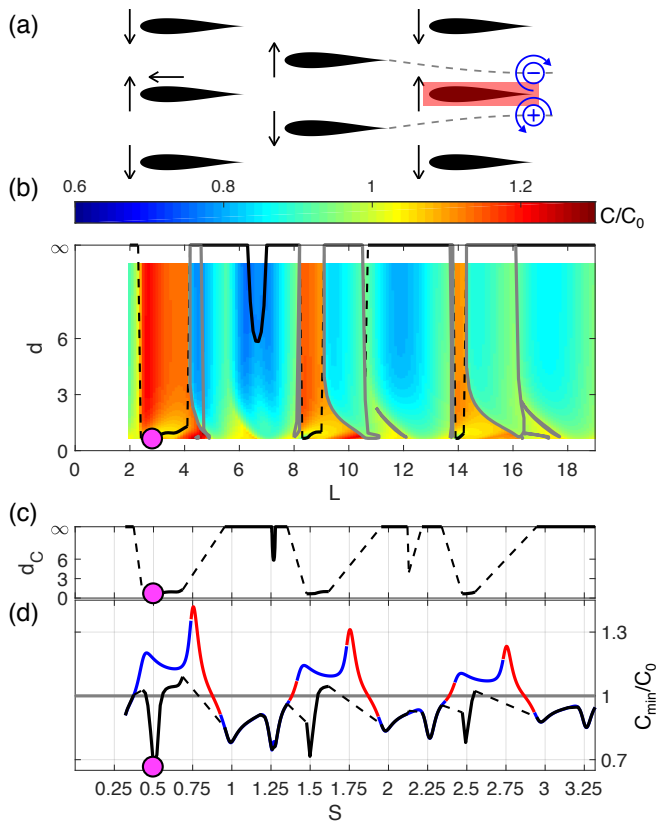}%{ArrayFigCOT-1.pdf}
      \caption{{\bf Cost of transport in a diamond lattice.} The procedure described in the caption of Fig.~\ref{Fig:Rect_COT} is repeated for a diamond lattice. The state with the lowest cost of transport has $C/C_0 = 0.67$ for $L = 2.8$ and $d = d_{\text{min}}$, corresponding to $S = 0.5$. %(a) Schematic of the diamond lattice of swimmers that minimizes the cost of transport $C/C_0$.  (b) Normalized time-averaged cost of transport $C_{\text{av}}/C_0$ as a function of $L$ and $d$, based on numerical simulations of the diamond lattice. The black curve $d^*_C(L)$ indicates the optimal lateral spacing as a function of the streamwise spacing $L$. The minimal COT of $C/C_0=0.67$ occurs at $L = 2.8$, $d = d_{\text{min}}$, corresponding to $S = 0.5$. (c) Optimal lateral spacing $d^*_C(L)$ plotted as a function of the schooling number $S$. (d) The associated cost of transport $C_{\text{min}}/C_0$ (black), superimposed on top of the results for a linear school from Fig.~\ref{Fig:LineSpeed}c.
 }
      \label{Fig:Array_COT}
\end{figure}

The complete equations are detailed in Supplemental Material Eq. (45). %\eqref{ModelDiamond}. 
Since the lattice is no longer symmetric about the midplane $y=0$, the steady state $u_n = U$ is not a solution to the governing equations, %, which is from inspection of the vortex arrangement in Fig.~\ref{Fig:Array}b and the corresponding arrangement after one time step. Instead, 
but the period-2 trajectory $u_n = U+(-1)^nU_1$ is. %a solution to the governing equations, where 
The unknowns $U$ and $U_1$ satisfy the pair of algebraic equations 
\footnotesize
\begin{align}
&\phantom{}\pm 2U_1 + F_D(U\pm U_1)=F_0\nonumber \\
&\phantom{}+\sum_{j=-2}^2\sum_{m=0}^{\infty}G\left(U\left(m+\frac{1}{2}\right)\pm\frac{U_1}{2}-\frac{jL}{2}\mp	\alpha_jd,U\pm U_1,m\right),\label{Diamond_BS}
\end{align}
\normalsize
where $\alpha_j = \text{mod}(j,2)/2$ accounts for the vertical shift in the columns at $x = \pm L/2$. This equation describes the balance of drag and vorticity-induced thrust on a swimmer, both of which depend on the swimmer's instantaneous flapping phase.  We describe how to solve these equations numerically, and extend the methodology described in \S\ref{Sec:1DStab} to assess the stability of these period--2 solutions, in Supplemental Material \S{}V %\ref{App:2D} 
and \S{}VI, %\ref{App:2DStab}, 
respectively. %Specifically, we show that the eigenvalues of the linear stability problem are given by the complex roots of the function $\mathcal{F}(z)$ in Eq.~\eqref{2DFeqn}, and we locate these roots numerically~\cite{Delves_1966}. Figure~\ref{Fig:ArraySoln} shows plots of these period--2 solutions as a function of $d$ for two different values of $L$, color-coded according to their stability. Note that solution branches can either be annihilated (Fig.~\ref{Fig:ArraySoln}a) or created (Fig.~\ref{Fig:ArraySoln}b) as the vertical spacing between swimmers is progressively decreased ($d\searrow d_{\text{min}}$), depending on the value of $L$. While the solution structure is evidently complex, we typically find that the \lq\lq fast mode\rq\rq\, survives when solution branches merge, as shown in Fig.~\ref{Fig:ArraySoln}a. This suggests that hydrodynamic interactions in a 2D geometry can enhance the school's speed.

We then numerically simulate the governing equations for a diamond lattice with initial conditions determined by the period--2 base states. As with the rectangular lattice described in \S\ref{Sec:Rect}, the simulations reveal the existence of multiple coexisting states, indicated by the regions of parameter space bounded by the gray curves in Fig.~\ref{Fig:Array}b. In such cases, Fig.~\ref{Fig:Array}b shows the normalized average speedup $U_{\text{av}}/U_0$ of the fastest state. We find that the formation experiences a maximum speedup of 22\% for $L = 4.4$ and $d = d_{\text{min}}$, which corresponds to $S = 0.67$. As with the rectangular lattice considered in \S\ref{Sec:Rect}, $d_U(L)$ is not identically equal to $d_{\text{min}}$, although the lateral spacing is indeed minimized for the fastest lattice. 

These results may also be qualitatively understood using a simple argument based on fluid flows, and by comparing the diamond lattice and in-line formation in Fig.~\ref{Fig:Array}d. As with the rectangular lattice, the slow modes ($U/U_0 < 1$) for an in-line formation exhibit speedup in a diamond lattice with the minimum lateral spacing, $d_U(L) = d_{\text{min}}$. Most fast modes for an in-line formation ($U/U_0 > 1$) benefit minimally from the diamond geometry; however, those with $S\lesssim s + 3/4$ experience a speedup of 3--5\% relative to the in-line formation, also with $d = d_{\text{min}}$. This is due to the existence of an entirely new branch of fast modes that emerges for $d\gtrsim d_{\text{min}}$, as shown in Supplemental Fig. 3. %\ref{Fig:L4p4_branch}. 
These fast modes may be attributed to the beneficial dipolar structure generated by the swimmer's upstream neighbors at $z = -L/2\pm id/2$, as shown in Fig.~\ref{Fig:Array}a. The fluid flow induced by this dipole at the midplane $y=0$ does not have a vertical component, and its horizontal component is negative, effectively imparting additional thrust to the swimmer and increasing its speed. The next flap generates a dipole of opposite sign, which imparts a drag; however, this contribution is weaker than the previous thrust contribution, since the associated vortices are farther from the body. 

Figure~\ref{Fig:Array_COT} shows the corresponding results for the cost of transport of a diamond lattice of swimmers. The cost of transport assumes the minimum value $C/C_0 = 0.67$ for a state with $S = 0.5$ and $d = d_{\text{min}}$, indicating that a tightly packed diamond lattice formation realizes a substantial increase in efficiency due to hydrodynamic interactions. Figure~\ref{Fig:Array_COT}b shows that, for a given streamwise spacing $L$, the optimal lateral spacing is typically $d_C(L)\approx d_{\text{min}}$ or $d_C(L)=\infty$. This effect is similar to that observed for the rectangular lattice, and may be rationalized by the physical argument presented in \S\ref{Sec:Rect}. However, the diamond lattices for which $S\approx s + 1/2$ are noticeably more efficient than their in-line formation counterparts, an effect that is absent for rectangular lattices. This finding may also be explained by the beneficial dipolar structure generated by the swimmer's upstream neighbors at $z = -L/2\pm id/2$ (Fig.~\ref{Fig:Array_COT}a), which provides an additional lift force as the swimmer accelerates upward.

%The diamond lattice evidently outperforms the rectangular lattice, phalanx and linear schools. We then investigated slight perturbations to this geometry by repeating our analysis for irregular lattices, in which the columns at $x = \pm L/2$ are shifted vertically by $0.05d$. As shown in Appendix Fig.~\ref{Fig:ArrayShift}, a downward shift yields a maximum speedup of $21\%$, and the upward shift a speedup of 20\%. Both optima occur at $S\approx 0.69$ and\footnote{The value $d_{\text{min}}\equiv (h_0+d_{\text{f}})/0.45$ is selected by the requirement that the upstream vortices do not pass through the swimmer, a phenomenon that cannot be modeled within our framework.} $d = d_{\text{min}}$. Taken together, these results suggest that the diamond lattice might be the geometry for realizing the maximum schooling speed, and that the hydrodynamic advantage is robust to small perturbations of this geometry.
%\\

While the distinct optimal diamond lattice formations shown in Fig.~\ref{Fig:Array}a and Fig.~\ref{Fig:Array_COT}a correspond to stable period-2 states, closer examination of the simulated solutions reveals that they often exhibit a complex nonlinear dynamics. This finding indicates that hydrodynamic interactions may significantly influence schooling behavior when the swimmers are interacting strongly,  particularly in the regime $d\gtrsim d_{\text{min}}$. We note that we also observed chaotic solutions in our model for the in-line formation, and leave the complete characterization of the system's nonlinear dynamics for future work.
%Indeed, we find that the schooling state with the maximum speedup $U_{\text{av}}/U_0$ in Fig.~\ref{Fig:Array}d is not a period--2 solution, but is instead temporally aperiodic. Figure~\ref{Fig:ArraySoln}b (inset) shows that the system may exhibit a period-doubling route to chaos as the vertical spacing $d$ is progressively decreased. (1D: see flip $h_0 =2$ cm branch at large $f$).

\section{Discussion}

We have presented a new model for the hydrodynamic interactions between swimmers in high-Reynolds number flows. While numerical simulations of such systems are computationally challenging, our model offers a simple framework by which to interpret the formation's dynamics: the swimmers shed vortices during each stroke, and in turn are propelled due to the vorticity-induced flow field. Despite neglecting consideration of the details of the flapping kinematics and the associated flow structures, our model exhibits good agreement with experimental data on interacting wings~\cite{Becker}, while using only three fitting parameters ($C_D$, $\tau$ and $C_v$). As shown in Fig.~\ref{Fig:data}, the observed bistability of slow and fast states is a consequence of overlapping branches of stable (blue) steady-state solutions, which are separated by unstable (red) branches. The multi-stability of states has not been observed in prior theoretical investigations, but is a generic feature of our model:  indeed, multiple coexisting states are found in the in-line, rectangular and diamond lattice formations for appropriate values of the geometric parameters. Our model also predicts new schooling instabilities (Supplemental Fig. 2) %\ref{Fig:dynamics}) 
through which the speed oscillates in time, an effect that can be probed experimentally. Animal schools might employ active control mechanisms in order to mitigate the effects of such instabilities.

%Since our model for a linear school is validated against experimental data, it gives insight into the lattice formations that allow for the largest schooling speed. 
Despite the apparent complexity of the hydrodynamic interactions, we show that the interaction thrust force between two swimmers is approximately proportional to the vertical velocity of a swimmer and the vertical velocity of the flow induced by its neighbor (Eq.~\eqref{FApprox}). This shows that the speed of an in-line formation is sensitive to the streamwise spacing, as a self-propelling flapping swimmer generates a spatially oscillatory flow field in its wake (Fig.~\ref{Fig:LineSpeed}d). %, obtaining speed increases for $n+1/4\lesssim S\lesssim n+3/4$ and decreases for $n+3/4\lesssim S\lesssim n+5/4$. 
For the set of flapping kinematics considered, we find that an in-line formation may move up to 17\% faster than an isolated swimmer, provided the distance between swimmers $L$ is such that the schooling number $S\lesssim 0.75$ (Fig.~\ref{Fig:LineSpeed}b). By contrast, a phalanx formation of swimmers flapping in antiphase moves roughly 5\% faster than a single swimmer (Fig.~\ref{Fig:Phalanx2}), with the optimal speedup occurring when the lateral distance between swimmers is minimized. While the fastest rectangular lattice provides a marginal advantage over an in-line formation (Fig.~\ref{Fig:Rect}), the fastest diamond lattice is able to move 22\% faster than a single swimmer (Fig.~\ref{Fig:Array}), provided that the lateral distance between swimmers is minimized. This effect may be attributed to the advantageous horizontal flow generated by a swimmer's upstream neighbors (Fig.~\ref{Fig:Array}a). %While the speedup in a phalanx formation is consistent with the recent experiments on schooling red nose tetra fish~\cite{AshrafPNAS}, our results suggest that swimmers could realize even larger speed increases by harnessing the fluid flow generated by their streamwise neighbors.

By using the cost of transport to measure the energetic efficiency, we find that in-line formations may realize an energy savings of 25\% over an isolated swimmer, provided the distance between swimmers $L$ is such that the schooling number $S\gtrsim 1.25$ (Fig.~\ref{Fig:LineSpeed}c). While our finding that the phalanx formation affords a speedup is in agreement with the experimental observations of Ashraf {\it et al.}~\cite{AshrafPNAS}, we also find such formations to be slightly less efficient than an isolated swimmer (Fig.~\ref{Fig:Phalanx2}). Generally, we observe that in-line and phalanx formations exhibit a tradeoff between speed and efficiency. While all rectangular lattices have a higher cost of transport than the most efficient in-line formation (Fig.~\ref{Fig:Rect_COT}), the most efficient diamond lattice has an energy savings of 33\% over an isolated swimmer, an effect that can also be attributed to the effective vortex dipole generated by the swimmer's upstream neighbors (Fig.~\ref{Fig:Array_COT}a). Taken together, our results show that the fastest and most efficient diamond lattice formations are distinct, and that they outperform the other geometries in speed and efficiency, respectively.   \markblue{Based on the argument presented at the end of \S\ref{Sec:Line}, we expect   qualitatively similar results for a model in which the swimmers flap perfectly out-of-phase with respect to their streamwise neighbors in both in-line and 2D lattice formations. %This is because both the speed and efficiency of 2D lattices are largely determined by the streamwise interactions between swimmers. 
While most of the relevant schooling numbers will be shifted, $S\rightarrow S+1/2$, we still expect the diamond lattices with $S = s+1/2$ to be particularly efficient (Fig.~\ref{Fig:Array_COT}d), owing to the vortex dipole shed by a swimmer's upstream neighbors.}

Our finding that the most efficient state is realized by a diamond lattice with the minimal lateral spacing $d = d_{\text{min}}$ is consistent with the results of Weihs~\cite{Weihs1,Weihs2}. However, our results differ in that the curves $d_U(L)$ and $d_C(L)$ (black curves in Fig.~\ref{Fig:Array}d and Fig.~\ref{Fig:Array_COT}d, respectively) are not monotonic, implying that there is not an optimal lattice angle. %, as claimed by Weihs~\cite{Weihs2}. 
While Weihs argued that it is disadvantageous for a swimmer to swim directly behind another, we find that this is not necessarily the case (Fig.~\ref{Fig:LineSpeed}). We note that our model differs from Weihs' in some important ways. First, we model flapping swimmers, %while Weihs neglects the swimmers' kinematics entirely. 
and find that the formation's speed is influenced by the relationship between the swimmer's kinematics and oncoming fluid flow (\S\ref{Sec:Line}). More sophisticated computational models~\cite{Maertens_Optim,Verma_RL} of fish-like swimming have also highlighted the importance of this relationship. Second, we allow the vortex strength to decay in our model, which sets an effective interaction distance $\sim U\tau$ between swimmers. Third, we explicitly account for the formation's dynamics and thus the schooling modes' stability, which was not possible in Weihs' framework.  %Our model predicts the emergence of a chaotic schooling mode in a 2D diamond lattice when $d\gtrsim d_{\text{min}}$ (Fig.~\ref{Fig:ArraySoln}b inset), indicating that schooling swimmers might need to use active control mechanisms in order to maintain a constant swimming speed. 

Our results may also be compared with recent theoretical and computational studies of a swimmer in a doubly periodic domain, which generates a lattice of swimmers flapping in phase. Recently, Hemelrijk {\it et al.}~\cite{Hemelrijk} conducted 2D numerical simulations using a multi-particle collision dynamics model and also found that the diamond lattice provides the largest speedup relative to a single swimmer ($23\%$), but that the maximum occurred for $d=2$, the {\it largest} lateral spacing considered. Similarly, the diamond lattice for $d = 1.75$ was found to have the highest Froude efficiency, or the ratio of the useful power to the total power input. However, Hemelrijk {\it et al.}~\cite{Hemelrijk} did not consider the influence of the streamwise spacing $L$, which we find to play a dominant role. Daghooghi \& Borazjani~\cite{Daghooghi} conducted 3D large eddy simulations of rectangular lattices of swimmers at high Reynolds number. They found that the swimming speed and efficiency increase as the lateral distance $d$ decreases, but also did not consider the influence of the streamwise spacing and instead fixed $L = 1$. They attributed the observed hydrodynamic advantage to channeling or wake blockage; this effect is beyond the scope of our model, since we assume the vortices to remain in place once shed. Nevertheless, in our simulations of rectangular lattices of swimmers that flap in antiphase with respect to their lateral neighbors, we find that the swimming speed and efficiency do not necessarily exhibit a monotonic dependence on $d$, but instead are nontrivially influenced by both $L$ and $d$ (Fig.~\ref{Fig:Rect}b and Fig.~\ref{Fig:Rect_COT}b). Tsang \& Kanso~\cite{Kanso_Dipole} proposed a far-field hydrodynamic model of swimmers as finite-sized vortex dipoles, and found that swimmers in both rectangular and diamond lattices actually move {\it slower} than they would in isolation. They attributed this result to the absence of shed vorticity in their model, which has been shown to mediate the near-field interactions between swimmers~\cite{Sophie}. Our dynamical model builds on these studies by explicitly modeling both the flapping kinematics and shedding of vortices. The model's mathematical simplicity allows us to analytically show the multi-stability of states in in-line, rectangular and diamond lattice formations, a result absent from all of the studies described above. Moreover, the model's computational tractability allows us to conclusively determine the dependence of the speed and cost of transport on the geometric parameters $L$ and $d$.%A similar model, coupled with behavioral interaction rules, is proposed by Gazzola {\it et al.}~\cite{Gazzola_JFM}, who find that the optimal rectangular school is to pack in traveling direction and separate far laterally.

\markblue{We now compare some of our results to schooling formations reported in the literature, despite the current sparsity of quantitative data. In their observations of red nose tetra fish, Ashraf {\it et al.}~\cite{AshrafPNAS,AshrafINT} reported that high-speed schools typically adopt phalanx formations with nearest neighbors separated by 0.5--0.6 body lengths. This observation is consistent with our theoretical prediction that the fastest phalanx formation is realized for the minimum lateral distance considered, $d\approx 0.7$ body lengths (Fig.~\ref{Fig:Phalanx2}). Similarly, Atlantic bluefin tuna have been observed to adopt phalanx formations with an average lateral spacing between 1 and 1.5 body lengths~\cite{Newlands_tuna,Partridge2}. When in relatively large schools, this fish species has also been observed to adopt diamond lattice-like formations with a mean first- and second-nearest neighbor separation angle between $14^{\circ}$ and $17^{\circ}$ (Fig. 5 in~\cite{Partridge2}). This observation is roughly consistent with our theoretical prediction that the diamond lattice with the lowest cost of transport has a separation angle of 13$^{\circ}$ (Fig.~\ref{Fig:Array_COT}). However, the red nose tetra fish has been observed to adopt diamond lattice formations with a typical separation angle of approximately $ 37^{\circ}$ (Fig. 2b in~\cite{AshrafPNAS}). Moreover, this fish species tends to adopt the phalanx over the diamond lattice formation at high swimming speeds~\cite{AshrafPNAS}, an observation that runs counter to our prediction that the fastest and most efficient states are realized by diamond lattice formations. We note that the spatial phase synchronization between neighboring fish, as measured through the schooling number $S$, is not typically reported, which prevents further quantitative comparison between our theoretical predictions and field observations. Such detailed comparison against biological swimmers would also benefit from more accurate modeling of the swimming kinematics and body shape. \markred{Considerations beyond hydrodynamics, such as social cues and predator avoidance, undoubtedly also impact the structure of schools observed in nature.}   %Partridge tuna is actually more precise: parabolic school has NND 0.8 to 1.3 BL. Newlands finds mean body length separation has modal values at 0.3 and 0.9 BL across all formations, whereas yours has 2.9. Newlands dominant angle also contradicts Partridge2 (tuna); she says 30-deg, despite same species! Partridge2 (tuna) says that across all schools get more like 1.5 to 2 body lengths NND distance. 
}

We expect the results presented herein to be most relevant for understanding fish schools, since we neglected the influence of lift forces on the dynamics. \markblue{While a recent study of shorebird flocks found no evidence of temporal or spatial phase synchronization between birds~\cite{Corcoran_V}, ibis flocks were observed to preferentially assume V-formations with median schooling number $S\approx s+1/4$, and in-line formations with $S\approx s+1/2$ (Supplemental Fig. 2 in~\cite{Portugal}). Extensions of our model might shed light on these phenomena, for which lift generation is an important consideration~\cite{Higdon}. We also note that a conceptually similar iterated map model may readily be applied to 3D flocks and schools; however, new techniques would be required to capture fully 3D dynamics, since the complex-variable techniques used herein cannot simply be extended to calculate the vortex-induced flow and associated forces on the bodies.}  Our work may also have limited application to \lq\lq disordered\rq\rq\, schools and cluster flocks, which deviate significantly from ordered lattice formations. The model may be generalized to allow for more general body kinematics, including pitching, turning and adaptive spacing between swimmers~\cite{Sophie,Newbolt}. %Koumotsakos~\cite{Novati_Learning} modify gait adaptively -- extend? More sophisticated model for shed vorticity, spatiotemporal vortex decay and drag laws could improve the accuracy.

While the quantitative results presented in this paper depend on the model parameters used, our reduced modeling framework may be applied to understand temporally long-lived hydrodynamic interactions in active systems. Indeed, models for the pilot-wave dynamics of droplets bouncing on a vibrating fluid bath have a similar mathematical structure~\cite{annualReview}, and may be extended to probe the droplet's complex collective dynamics~\cite{Lieber-lattices,lattices}. % cut ref to lattice2
Generally, we expect models of the type described herein to be broadly applicable to systems of active particles interacting via their collective histories. %Possible extension to turbines~\cite{Dabiri_Turbine}?

\acknowledgments{We thank Hassan Masoud, Fang Fang, Joel Newbolt, Sophie Ramananarivo and Stephen Childress for helpful discussions. A. U. O. acknowledges support from the Simons Foundation (Collaboration Grant for Mathematicians, award \#587006), L. R. acknowledges support from the NSF (DMS-1847955), and M. J. S. thanks the Lilian and George Lyttle Chair of Applied Mathematics. %AO gratefully acknowledges the support of the NSF Mathematical Sciences Postdoctoral Fellowship.
}

%\section*{References}
%\nocite{*}
%\bibliographystyle{apsrev4-1}

%\bibliographystyle{abbrv}
\bibliographystyle{unsrtnat}
\bibliography{SchoolingBib,biblio-Oza,DropReferences}

\end{document}